\documentclass[fleqn,usenatbib]{mnras}

\usepackage{mathptmx}

\usepackage[T1]{fontenc}
\usepackage{ae,aecompl}

\usepackage{graphicx}	
\usepackage{amsmath}	
\usepackage{amssymb}	
\usepackage{epstopdf}
\usepackage{textcomp}

\title[Dynamical evolution and stability maps of the Proxima Centauri system]{Dynamical evolution and stability maps of the Proxima
Centauri system}
\author[Tong Meng et al.]{Tong Meng$^{1,2}$, Jianghui Ji$^{1}$\thanks{E-mail: jijh@pmo.ac.cn}, Yao Dong$^{1}$\\
$^{1}$CAS Key Laboratory of Planetary Sciences, Purple Mountain Observatory, Chinese Academy of Sciences, Nanjing 210008, China\\
$^{2}$University of Chinese Academy of Sciences, Beijing 100049, China\\
}

\begin{document}

\date{Received 2012 July 26; in original form 2011 October 30}

\pagerange{\pageref{firstpage}--\pageref{lastpage}} \pubyear{2012}

\maketitle

\label{firstpage}

\begin{abstract}
  Proxima Centauri was recently discovered to host an Earth-mass planet of Proxima b, and a 215-day signal which is probably a potential planet c.
  In this work, we investigate the dynamical evolution of the Proxima Centauri system with the full equations of motion and semi-analytical models including relativistic and tidal effects. We adopt the modified Lagrange-Laplace secular equations to study the evolution of eccentricity of Proxima b, and find that the outcomes are consistent with those from the numerical simulations. The simulations show that relativistic effects have an influence on the evolution of eccentricities of planetary orbits, whereas tidal effects primarily affects the eccentricity of Proxima b over long timescale. Moreover, using the MEGNO (the Mean Exponential Growth factor of Nearby Orbits) technique, we place dynamical constraints on orbital parameters that result in stable or quasi-periodic motions for coplanar and non-coplanar configurations. In the coplanar case, we find that the orbit of Proxima b is stable for the semi-major axis ranging from 0.02 au to 0.1 au and the eccentricity being less than 0.4. This is where the best-fitting parameters for Proxima b exactly fall.
  Additional simulations show that the robust stability of this system would favor an eccentricity of Proxima b
  less than 0.45 and that of Proxima c below 0.65. In the non-coplanar case, we find that mutual
  inclinations of two planets must be  lower than $50^{\circ}$ in order to provide stability.
  Finally, we estimate the mass of Proxima c to be $3.13~{M_ \oplus } \le {m_c} \le 70.7~{M_ \oplus }$ when $1.27~{M_ \oplus } \le {m_b} \le 1.6~{M_ \oplus }$, if ${i_{mutual}} \le {50^{\circ}}$ and $\Delta \Omega  = 0^{\circ}$.
\end{abstract}


\section{Introduction}

At a distance of 1.295 pc, the red dwarf Proxima Centauri
($\alpha$ Centauri C or hereafter Proxima) is the Sun's
closest stellar neighbour. As one of the best-studied low-mass
stars, Proxima Centauri, accompanied by $\alpha$ Centauri AB, belongs to a triple-star system and has a mass of 0.12 $M_{\odot}$.
Proxima b was discovered recently to orbit Proxima Centauri \citep{Anglada-Escud¨¦2016}, with a minimum mass of 1.27 $M_ \oplus$,
an orbital period of 11.2 days, a semi-major axis of approximately 0.049 au, and an eccentricity
below 0.35. Moreover, a 215-day planet of Proxima c was suspected to exist and
revolve around Proxima Centauri, although this object may be an artifact arising from stellar activity combined with very uneven sampling \citep{Kurster2003,Anglada-Escud¨¦2016,Rajpaul2016}.

The discovery of Proxima b provides new clues to understand habitable planets \citep{Kasting1993,Kasting2003,Kopparapu2013}. Hence, it is essential to explore the physical parameters, dynamical evolution and atmospheric environment of Proxima b to better understand potential active biology \citep{Barnes2016}, by performing numerical simulations of the Proxima Centauri system to explore various planetary configurations. Moreover, different models were established to constrain the mass and radius, and compositions of Proxima b.
In a model of geomagnetic properties, \citet{Zuluaga2018} showed that Proxima b would be a terrestrial planet with a mass $1.3~{M_ \oplus } \le {M_p} \le 2.3~{M_ \oplus }$ and a radius ${R_p} = 1.4_{ - 0.2}^{ + 0.3}~{R_ \oplus }$ at the 70$\%$ confidence level. Recently, \citet{Bixel2017} indicated that the planet's density is in good accordance with the composition of a rocky planet at 95$\%$ confidence level of ${\left\langle M \right\rangle _{rocky}} = 1.63_{ -
0.72}^{ + 1.66}~{M_ \oplus }$ for its mass and ${\left\langle R \right\rangle _{rocky}} = 1.07_{ - 0.31}^{ + 0.38}~{R_ \oplus
}$ for its radius, rather than composed of ice materials or an H/He envelope. \citet{Brugger2017} gave an update on mass-radius relationship and offered an estimate of the planet's composition from density measurements where
its radius would reach 1.94 $R_{\oplus}$ for a 5 $M_{\oplus}$ planet, thereby concluding that the mass of Proxima b should be below this value at a 96.7$\%$ confidence level. Furthermore, the influence of orbital inclination on Proxima b was explored on the basis of the planetary mass and relevant physical properties, and the dynamical simulations indicated that the presence of additional terrestrial planets within the Habitable Zone changes as a function of inclination \citep{Kane2017}.
The capability of Proxima b to retain liquid water on the surface is related to the strong stellar irradiation \citep{Ribas2016}. However, \citet{Airapetian2017} recently indicated that the planet cannot be habitable, because an Earth-like atmosphere would escape within $\sim$10 Myr \citep{Jin14,Owen16,DongC2017,Jin18} due to the star's strong XUV flux. If Proxima b was catalogued as a super-Earth with a mass 5 $M_{\oplus}$, the escape velocity at the planet's surface would not be sufficient to hold the atmosphere \citep{Airapetian2017}. Finally, the habitability for rocky 'waterworld' planets is strongly affected by ocean chemistry \citep{Kite2018}. The investigations offer hints that Proxima b is more likely to be a terrestrial planet.

Current  planetary formation theory suggests that the gas drag in the
protoplanetary gaseous nebular may drive the planets to migrate into the place near 0.1 au away from the star \citep{Lin1996,Lee2002,Wang2012,Wang2014,Wang2017}. For close-in orbits, tidal effects may have a significant impact on the planet, such as tidal locking  \citep{Kasting1993} or circularizing the orbit \citep{Darwin1880,FerrazMello2008}. The interplay between planets and their host stars, e.g., tides, plays a key role in the dynamical evolution inside 0.1 au \citep{Nagasawa2008,Jackson2009}. In addition, general relativity
will further alter the orbit \citep{Mardling2007}. The equations of motion of the planets under mutual interaction were given in coplanar systems under the combined effects of general relativity and tides from the host star \citep{Mignard1979,Mardling2002,Beutler2005,Rodr¨ªguez2011,Dong2013,Dong2017}. Therefore, we should take into account these effects in the models when studying dynamical evolution of the Proxima Centauri system.

In our simulations, we first modified MERCURY6 package by adding the terms of relativistic and tidal effects \citep{Chambers1999, Dong2013}. Consequently, the modified package can be employed to explore the influence of relativity and tides on the planetary orbits over long timescale integrations \citep{Mignard1979, Beutler2005, Mardling2002, Rodr¨ªguez2011,Dong2013,Dong2017}. On the other hand, from a theoretical viewpoint, we modified the Lagrange-Laplace secular equations by incorporating the major terms of relativistic and tidal effects \citep{Laskar2012,Dong2014} to simulate the orbital evolution of two planets, although it is an approximation to the real motions of the Proxima Centauri system. The other major objective is to compare the semi-analytical results from the modified Lagrange-Laplace secular equations with the outcomes from direct numerical integrations of the system under study.

To better understand the stability of the system, we adopt MEGNO \citep{Cincotta2000} to explore a large number of system parameters. MEGNO is generally a good indicator to distinguish between the regular or chaotic orbits in planetary systems \citep{Cincotta2000,Nunez2000,Giordano2001}. The sets of orbital parameters that support stable (quasi-periodic) motions within planetary systems can be explored efficiently with MEGNO. For example, stability maps of the GJ 876 system constructed in the surrounding regions of the Laplace resonance \citep{Lee2002,Wang2012,Wang2014,Marti2016,Gozdziewski2016,Sun2017}. \citet{Gozdziewski2002} indicated that the ranges of the orbital elements that provide regular evolutions of the 47 UMa system. \citet{Gozdziewski2003a} placed the orbital parameter bounds and the impact of mass for the outer planet on the bounds of HD 37124 system using MEGNO stability maps. Moreover, \citet{Gozdziewski2003b} showed that the HD 12661 system evolves at a border of the 11:2 mean motion resonance, which remains stable and gives rise to quasi-periodic motion. In this work, we will utilize MEGNO to constrain the dynamical limits on orbital parameters that generate stable or quasi-periodic motions of the Proxima Centauri system for coplanar and non-coplanar orbits.

This paper is structured as follows:
in Section 2 we describe the modified Lagrange-Laplace secular equations and the full equations of motion
adopted in numerical simulations, which are supplemented with relativity and tides,
for the two-planet system of Proxima Centauri.
In Section 3, we explore the long-term dynamical evolution of the Proxima Centauri planets by
comparing the numerical results with those given in semi-analytical models.
In Section 4, we present the MEGNO stability map of Proxima Centauri system, and
further evaluate dynamical constraints on the parameter space
that can provide stable motions.
Finally, we summarize our conclusion and give a concise discussion in Section 5.

\section{Dynamical model}

\subsection{Modified secular perturbation theory}

General secular perturbation theory does not take into account relativistic and tidal effects
raised by the host star, but considers only Newtonian gravitational forces.
In the secular approximation, the semi-major axes remain constant, indicating that there is no exchange of energy
among the orbits. Only an exchange of angular momentum that causes variations in the eccentricities,
which are usually described by the classical Lagrange-Laplace equations \citep{Laskar1990,Murray1999,Ji2003,Ji2007}. Moreover,
the secular equations of motion augmented with relativistic and tidal effects are widely adopted to investigate
dynamical evolution especially for short-period planets \citep{Eggleton2001,FerrazMello2008,Correia2011,Laskar2012}.
Here, we focus on coplanar systems. We  define  $l$ as the number of planets. Using the classical
complex variables, ${z_k} = {e_k}{e^{i{w_k}}}$, for $k = 1,...,l$, the secular equations to first order in the
eccentricity are as follows \citep{Laskar2012}

 \begin{equation}
 \frac{{d\mathbf{z}}}{{dt}} = i\mathbf{A}\mathbf{z}\
 \end{equation}
 where $\mathbf{z} = {\left[ {{z_1},{z_2},...,{z_l}} \right]^T}$, $\mathbf{A}$ is a real matrix whose elements are \citep{Laskar1995}
 \begin{equation}
 {A_{jj}} = \sum\limits_{k = 1}^{j - 1} {{n_j}} \frac{{{m_k}}}{{{m_0}}}{C_3}\left( {\frac{{{a_k}}}{{{a_j}}}} \right) + \sum\limits_{k = j + 1}^l {{n_j}} \frac{{{m_k}}}{{{m_0}}}\frac{{{a_j}}}{{{a_k}}}{C_3}\left( {\frac{{{a_j}}}{{{a_k}}}} \right)\
 \end{equation}
 and
 \begin{equation}
 {A_{jk}} = \left\{ \begin{array}{l}
 2{n_j}\frac{{{m_k}}}{{{m_0}}}\frac{{{a_j}}}{{{a_k}}}{C_2}\left( {\frac{{{a_j}}}{{{a_k}}}} \right)\;\quad \quad \quad \;j < k \\
 2{n_j}\frac{{{m_k}}}{{{m_0}}}{C_2}\left( {\frac{{{a_k}}}{{{a_j}}}} \right)\quad \quad \quad \quad \quad j > k \\
 \end{array} \right.\
 \end{equation}
 where ${m_k}$, ${a_k}$, ${n_k}$ are, respectively, the mass, the semi-major axis, the average angular velocity of
 the $k$th planet, the index 0 represents the star. Let $\alpha$  be the ratio of two semi-major axes,
 ${C_2}\left( \alpha  \right)$ and ${C_3}\left( \alpha\right)$ are functions of the Laplace
 coefficients \citep{Laskar1995,Murray1999}.

The above secular equations do not include relativistic and tidal effects. We thus define two new
diagonal matrices $\delta\mathbf{A}$ and $\delta\mathbf{B}$, and the full secular evolution
is given by  \citep{Laskar2012}

 \begin{equation}
 \frac{{d\mathbf{z}}}{{dt}} = \left( {i{\mathbf{A}_{tot}} - \delta\mathbf{B}} \right)\mathbf{z}\
 \end{equation}
 where ${\mathbf{A}_{tot}} = \mathbf{A} + \delta\mathbf{A}$, $\delta {A_{kk}} = \delta {A_{kk}}^{\left( 1
 \right)} + \delta {A_{kk}}^{\left( 2 \right)}$, the superscript 1 and 2 refer to
 relativistic and tidal effects respectively.
 The effect of relativity on the $k$th planet is conservative, and to first order in eccentricity is
 given by

 \begin{equation}
 \delta {A_{kk}}^{\left( 1 \right)} = 3\frac{{G{m_0}}}{{{c^2}}}\frac{{{n_k}}}{{{a_k}}}\
 \end{equation}
Tidal effects have two contributions on the conservative term and the dissipative part, ${\delta\mathbf{B}}$ represents the tidal
effect of the dissipative term. We have

 \begin{equation}
 \delta {A_{kk}}^{\left( 2 \right)} = \frac{{15}}{2}{K_k}\
 \end{equation}
 \begin{equation}
 \delta {B_{kk}}^{\left( 2 \right)} = 27\left( {1 - \frac{{11}}{{18}}\frac{{{w_k}}}{{{n_k}}}}
 \right)\frac{{{K_k}}}{{{Q_k}}}\
 \end{equation}
 where
 \begin{equation}
 {K_k} = {k_{2,k}}{n_k}\left( {\frac{{{m_0}}}{{{m_k}}}} \right){\left( {\frac{{{R_k}}}{{{a_k}}}} \right)^5}
 \end{equation}
 where ${R_k}$, ${w_k}$, ${k_{2,k}}$, ${Q_k}$ are, respectively, the radius, the proper rotation rate, the
 second Love number, the dissipation coefficient of the $k$th planet. Generally, the modified dissipation
 coefficient is adopted, ${Q_k}' = 3{Q_k}/(2{k_{2,k}})$. Finally, the solutions are

 \begin{equation}
 {u_k}\left( t \right) = {u_k}\left( 0 \right){e^{ - {\gamma _k}t}}{e^{i{g_k}t}}\
 \end{equation}
 where ${\gamma _k}$ are the coefficients arising from tidal dissipation and ${g_k}$ the eigenvalues of real
 matrix ${\mathbf{A}_{tot}}$. Applying it to a two-planet system, to first order, the eccentricity variables
 ${z_1}$ and ${z_2}$ are linear combinations of ${u_1}\left( t \right)$ and ${u_2}\left( t \right)$

 \begin{equation}
 {z_1}\left( t \right) = {S_{11}}{u_1}\left( t \right) + {S_{12}}{u_2}\left( t \right)\
 \end{equation}
 \begin{equation}
 {z_2}\left( t \right) = {S_{21}}{u_1}\left( t \right) + {S_{22}}{u_2}\left( t \right)\
 \end{equation}
 where ${S_{11}}$, ${S_{12}}$, ${S_{21}}$ and ${S_{22}}$ are the elements of a complex matrix
 $\mathbf{S}$.

\subsection{Numerical model}

In this section, we briefly introduce the model of our numerical
simulations for the Proxima Centauri system. Here the reference frame adopted
is centered at the host star and the planetary orbits are coplanar with
respect to the reference plane. We consider the relativistic and tidal
effects from the central star acting on the two planets.
The equations of motion are \citep{Rodr¨ªguez2011}

\begin{eqnarray}\label{mov}
\ddot{\mathbf{r}}_i&=&-\frac{G(m_0+m_i)}{r_i^3}\mathbf{r}_i+Gm_j\Bigg{(}\frac{\mathbf{r}_j-\mathbf{r}_i}{|\mathbf{r}_j-\mathbf{r}_i|^3}-\frac{\mathbf{r}_j}{r_j^3}\Bigg{)}\\
&&+\frac{(m_0+m_i)}{m_0m_i}(\mathbf{f}_{\textrm{t}i}+\mathbf{f}_{\textrm{g}i})+\frac{\mathbf{f}_{\textrm{t}j}+\mathbf{f}_{\textrm{g}j}}{m_0}
\end{eqnarray}
where $i,j = 1,2$ and $i \ne j$. Subscript 1 and subscript 2 denote Proxima b and Proxima
c, respectively. Subscript 0 represents the central star. $\mathbf{f}_{\mathrm{\textrm{g}1}}$ and $\mathbf{f}_{\mathrm{\textrm{g}2}}$
are the general relativity contributions to  the inner and the outer planet, respectively. They are approximated
by \citep{Beutler2005,Rodr¨ªguez2011}

\begin{equation}
{\mathbf{f}_\mathrm{gi}} = \frac{{G{m_0}{m_i}}}{{{c^2}{r_i}^3}}\left[ {\left( {4\frac{{G{m_0}}}{{{r_i}}} - {\mathbf{v}_i}^2}
\right){\mathbf{r}_i} + 4\left( {{\mathbf{r}_i} \cdot {\mathbf{v}_i}} \right){\mathbf{v}_i}} \right]\
\end{equation}
where $\mathbf{v}_i$ = $ \mathbf{\dot{r}}_i$ and $c$ is the speed of light. Additionally, $\mathbf{f}_{\mathrm{\textrm{t}i}}$ is the tidal force  exerted  on the $i$ body respectively. Accordingly we utilize
the modified form in the following equation \citep{Mignard1979,Mardling2002,Rodr¨ªguez2011}

\begin{equation}
{\mathbf{f}_\mathrm{ti}} =  - \frac{{9G{m_0}^2{R_i}^5}}{{2{Q'}_i{n_i}{r_i}^{10}}}\left[ {2{\mathbf{r}_i}\left( {{\mathbf{r}_i} \cdot
{\mathbf{v}_i}} \right) + {r_i}^2\left( {{\mathbf{r}_i} \times {\boldsymbol{\Omega}_i} + {\mathbf{v}_i}} \right)} \right]\
\end{equation}
where $R_i$ is the radius of a planet, $\boldsymbol{\Omega}_i$ is the
angular velocity of rotation, $Q^\prime_i$ defined
as the modified dissipation coefficient which absorb the Love number as
$Q^\prime_i \equiv 3Q_i/2k_i$, which are associated with time lags
between tidal interaction and the corresponding deformation for the
planets, $G$ is gravitational constant. For Proxima b and Proxima c, the
typical value $Q_i^\prime$ = 100 for Earth-like planets is adopted
in this work. We also ignore the tidal effects raised by planets on the central star.

\subsection{Initial setup}

The adopted orbital elements and physical parameters of two planets are shown in Table 1.
According to the observations \citep{Anglada-Escud¨¦2016}, the initial orbital elements
are assumed as follows:
$a_{b}$ = 0.0485 au and $a_{c}$ = 0.346 au are the semi-major axes of each planet, respectively,  whereas $e_{b}$ = 0.05 and $e_{c}$ = 0.1, are the eccentricities. $\lambda_{b,c} {\rm{ =}}\omega_{b,c}  + M_{b,c} = {110^\circ}$, $\omega_b=\omega_c  = {310^\circ}$ and $M_b =M_c ={160^\circ}$, where $\lambda_{b,c}$, $M_{b,c}$ and $\omega_{b,c}$, stand for the mean longitude, the mean anomaly and the argument of periastron of each planet, respectively. The radius of Proxima b is assumed to be $1.07~R_{\oplus}$ \citep{Bixel2017}.
The radius of Proxima c is evaluated according to the mass-radius relationship of $m_p/M_\oplus$=2.69($R_p/R_\oplus$)$^{0.93}$ \citep{Weiss2014}. The stellar mass and radius are 0.120 $M_{\odot}$ and 0.1414 $R_{\odot}$ respectively \citep{Anglada-Escud¨¦2016}.
The inclinations of planets are discussed in the following sections where the results are described.

\section{Results of dynamical evolution}

Herein we adopt the modified Lagrange-Laplace secular
equations \citep{Laskar2012} and the modified numerical model to investigate the orbital evolution of the eccentricities of the Proxima Centauri system  by considering the unconfirmed Proxima c \citep{Anglada-Escud¨¦2016, Barnes2016}.
We obtain the dominant outcomes for dynamical evolution and compare these results.
We focus on coplanar systems in this section, and we assume
that the planets are synchronisation since the timescale of synchronisation is exceedingly short as compared with that of tidal evolution.

\begin{table}
 \begin{minipage}{90mm}
  \caption{The adopted orbital elements and physical data for Proxima b and the putative planet of Proxima c in Proxima Centauri system.
  \citep{Anglada-Escud¨¦2016,Barnes2016,Bixel2017}}
   \begin{tabular}{llllll}
  \hline
  Planet & m~sin$i$($M_{\oplus}$) & $P$(day) & $R$($R_{\oplus}$) &  $a$(au) & $e$  \\
  \hline
  Proxima b    &   1.27  & 11.2  & 1.07 & 0.0485 &     <0.35       \\
  Proxima c    &   3.13  & 215   & 1.18 & 0.346  &     0.1         \\
\hline
\end{tabular}
\end{minipage}
\end{table}

\subsection{Results of Modified secular theory}

Proxima b and Proxima c are close to the central star with semi-major axes of approximately 0.0485 au and 0.346 au, respectively.
Therefore, the relativistic and tidal effects exerted by the host star will play a major role in reshaping their final planetary orbits, especially for short-period planets or extremely close-in inner planets \citep{Eggleton2001,FerrazMello2008,Correia2011,Laskar2012}.
Thus, we can employ the semi-analytical Lagrange-Laplace secular equations to investigate the eccentricity evolution \citep{Laskar2012}.
We concentrate on the coplanar configurations and the initial parameters from Table 1 \citep{Anglada-Escud¨¦2016, Barnes2016}.
To explore tidal effects, we adopt the typical value $Q_i^\prime$ = 100 ($Q_i$=20 and $k_i$=0.3) for terrestrial planets in this work.

Fig.1 shows the evolution of eccentricity of Proxima b and Proxima c in the framework of classical Lagrange-Laplace secular equations (cLL) (\emph{green curves}), cLL plus relativistic effect (\emph{red curves}), and cLL plus relativistic and tidal effects (\emph{blue curves}).  We observe that the eccentricity evolution in the fluctuating periods and the amplitudes from the results of cLL+relativity and cLL+relativity+tide perfectly overlap over 1 Myr. This suggests that tidal effects have a tiny influence on the evolution of eccentricities for this system over short timescale. However, the evolution results from two latter models apparently differ from that of cLL. Moreover, we notice that the amplitude of eccentricity of Proxima b is significantly reduced  by  roughly $33\%$. And the period of oscillation drops from 0.214 Myr to 0.143 Myr.
As for Proxima c, there is a smaller influence on the amplitude of eccentricity that falls by about $20\%$, and the oscillating period is reduced because of relativistic effects.  The outcomes further show that the variations of eccentricities of Proxima b and Proxima c agree within 1{\textperthousand} in the model of cLL+relativity+tide.
Qualitatively, the results of modified secular equations indicate that relativistic effects  are 
more significant than tidal effects in the Proxima Centauri system over a timescale of 1 Myr.

\begin{figure}
\includegraphics[scale=0.35]{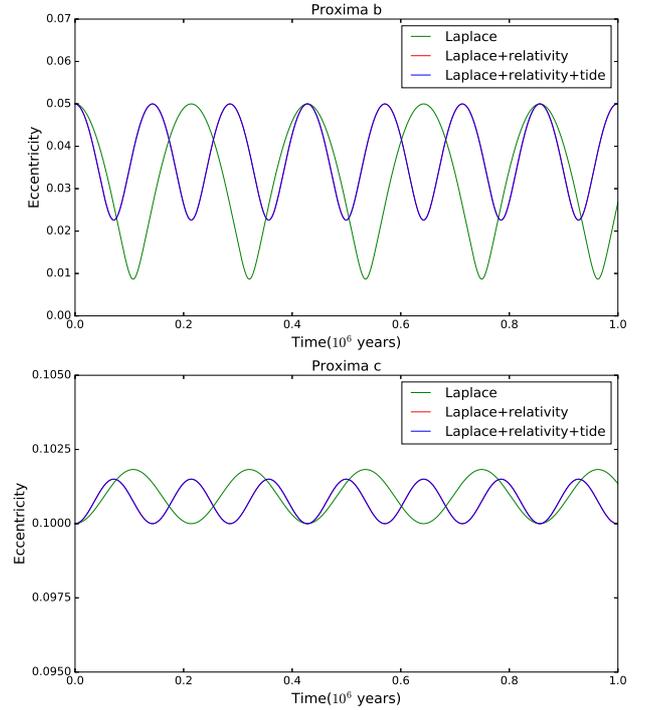}
  \centering
  \caption{The dynamical evolution of orbital eccentricities of the Proxima Centauri system within 1 Myr for coplanar cases, by taking into account the modified secular theory incorporated with relativistic and tidal effects. \textit{Upper Panel}: Proxima b.
  \textit{Bottom Panel}: Proxima c.
  The green lines, red lines and blue lines denote the evolution of eccentricities from cLL, cLL + relativity and cLL + relativity + tides, respectively.
  }\label{fig1}
 \end{figure}

\subsection{Results of Numerical model}

In this section, we mainly concentrate on the coplanar configuration and utilize the full equations of motion that include the relativistic and tidal effects \citep{Mignard1979,Beutler2005,Mardling2002,Rodr¨ªguez2011,Dong2013,Dong2017}, to investigate the eccentricity evolution of the planetary orbits.  We use the modified MERCURY6 code \citep{Dong2013} to perform numerical investigations. In our simulations, we use Bulirsch-Storer algorithm, in which the initial time step is 0.28 days (roughly 1/40 of the orbital period of the inner planet) and the accuracy parameter 10$^{-12}$, respectively.

For tidal effects, we set a typical value $Q_i^\prime$ = 100 ($Q_i$=20 and $k_i$=0.3) for terrestrial planets in this work.
In the simulations of Nbody (\emph{green curves})(see Fig.2), the maximum fractional energy change is $2.1 \times 10^{-6}$ and the largest fractional angular momentum change is $1.7 \times 10^{-7}$. In the case of Nbody+relativity (\emph{red curves}), the largest fractional energy variation is $2.2 \times 10^{-6}$ and the largest fractional angular momentum change is $ 1.8 \times 10^{-7}$. As shown in Fig.2, the cases with and without tide are almost coincident for the evolution of Proxima b and Proxima c, respectively, indicating that tidal effects have  a very small influence on the  evolution  of  the eccentricities. The simulation results show that amplitude of eccentricity of Proxima b is about 1{\textperthousand} due to tidal forces, whereas the contribution of tides  is even smaller for Proxima c.  Similar to the analytical model, comparing the results given by Nbody (\emph{green curves}) and Nbody+relativity+tide (\emph{blue curves}), we observe that the amplitude of eccentricity of Proxima b is significantly less by about $33\%$, whereas the oscillating period falls from 0.211 Myr to 0.141 Myr in the evolution.
For Proxima c, we find that the amplitude of eccentricity is also less by approximately $20\%$, whereas its oscillation period  drops from 0.211 Myr to 0.141 Myr.

Qualitatively, the numerical results  suggest  that relativistic effects are more important in the evolution than tidal effects does, while relativistic and tidal effects are more pronounced on inner planet (Proxima b) than outer planet (Proxima c). The results are consistent with those of the modified secular perturbation theory.

\begin{figure}
\includegraphics[scale=0.35]{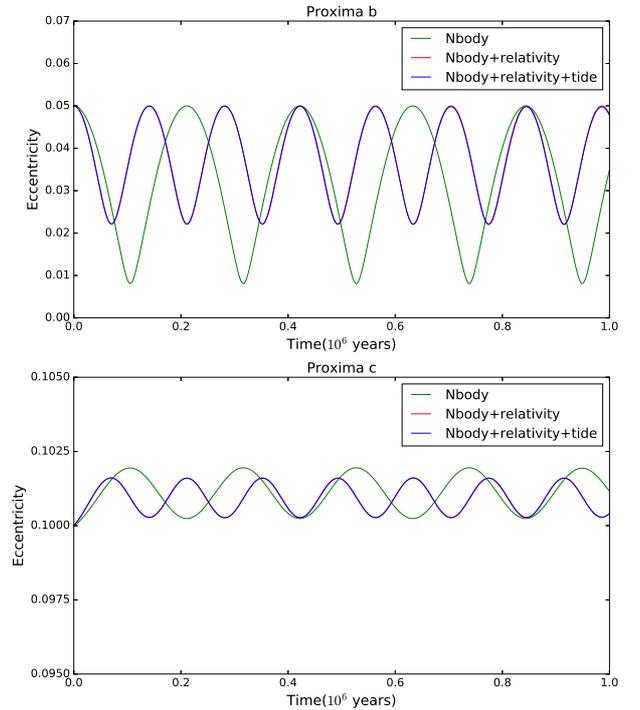}
  \centering
  \caption{The evolution of the orbital eccentricity of Proxima b and
  Proxima c within 1 Myr for coplanar cases, given by the numerical model including general relativity and tidal effect.
  The green lines, red lines and blue lines represent the evolution of eccentricities from the models of Nbody, NBody + relativity, and NBody + relativity + tide, respectively.
  }\label{fig2}
 \end{figure}

\subsection{Analysis of the dynamical evolution results}

Fig. 3 shows that the fluctuating periods of eccentricity of Proxima b are 0.143 Myr and 0.141 Myr, respectively, calculated from Laplace + relativity + tide (\textit{blue lines}) and NBody + relativity + tide (\textit{green lines}), which are in good agreement. We can draw a similar conclusion for the eccentricity evolution of Proxima c. Therefore, we can apply modified secular perturbation theory to investigate the evolution of eccentricities of two planets over much longer timescale.

Fig.4 displays the eccentricity evolution for the two planets over 7 Gyr with the modified secular theory.
There is no apparent orbital decay due to tides within the first million years,
but there are several minor deviations in the oscillations of the eccentricity of two planets due to relativistic effects.
However, the orbit of Proxima b gradually declines to finally be circularized over longer timescales as predicted by tidal model, whereas that of Proxima c remains an exceedingly tiny oscillation about 0.10, indicating that  tides have  no remarkable influence on its eccentricity evolution.
Our results are consistent with those by \citet{Barnes2016}, who pointed out that the eccentricity of Proxima b drops to $\sim$0.01 within 2-3 Gyr in a single planet system during tidal heating in the two-planet system \citep{Barnes2016}.
The Proxima Centauri system, which may host a potential planet of Proxima c, would resemble the Kepler 10 system,
as both systems bear one close-in inner planet and a distant outer companion \citep{Batalha2011}.
The investigation of dynamical evolution of Kepler 10 system showed that the orbit of inner planet can suffer tidal decay and circularization \citep{Dong2013} with the perturbation of the outer companion over long timescales, because Kepler 10b has a much closer orbit to its host star than that of Proxima b.

\begin{figure}
\includegraphics[scale=0.35]{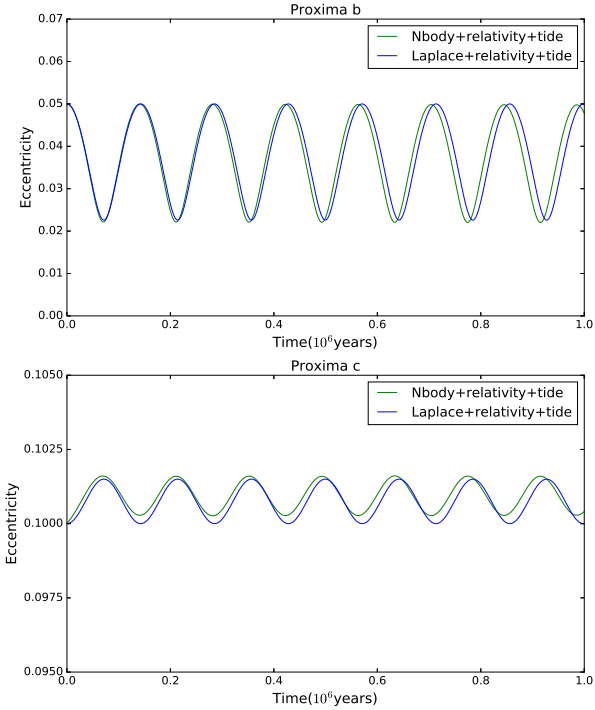}
  \centering
  \caption{The comparison of the evolution of orbital eccentricities for Proxima b and Proxima c for coplanar case within 1 Myr.
  The green curves represent the variation of eccentricity from numerical model of NBody + relativity + tide,
  whereas the blue profiles show  the results from Laplace + relativity + tide.
  }\label{fig3}
 \end{figure}

 \begin{figure}
\includegraphics[scale=0.45]{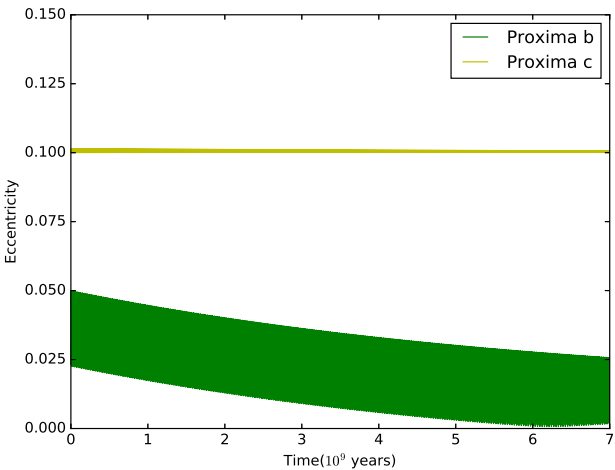}
  \centering
  \caption{The eccentricity evolution of Proxima b and c over timescale of 7 Gyr,
    using the modified secular perturbation theory considering general relativity and tidal effect.}
  \label{fig4}
 \end{figure}

\section{Stability maps of the Proxima Centauri system}

In this section, we employ MEGNO to place dynamical constraints on the stability of Proxima Centauri system
for coplanar and non-coplanar orbits.The MEGNO technique was invented by P. Cincotta and C. Sim{\'o}, without considering
the general relativity and tidal effects, to explore the dynamical behaviour of planetary systems
in the framework of the gravitational N-body problem. As mentioned-above, this method can rapidly distinguish between chaotic and regular evolution of a planetary system \citep{Cincotta2000,Nunez2000,Giordano2001,Gozdziewski2001}. Therefore, it can improve our understanding of the dynamical stability for this system.

\subsection{The MEGNO indicator}

Considering a planetary system as N point masses with gravitational interactions, we can characterize regular or
irregular states by computing the Lyapunov Characteristic Number (LCN) of the dynamical system.
However, MEGNO was further developed to distinguish between stable and chaotic orbits based on LCN.
By calculating LCN, we can obtain only one simple result that neglects dynamical information after long-term calculations, whereas identifying regular motion requires a time-consuming computation \citep{Murray1999}.
However, MEGNO is $10 - {10^2}$ times faster than
direct calculations of the LCE \citep{Cincotta2000,Nunez2000,Giordano2001,Gozdziewski2001}, which can be used to exhaustively  explore the parameter space for tens of thousands of initial conditions.

In this work, we consider the coplanar and non-coplanar cases. By setting a large number of
initial conditions, and then produce stability maps to identify the dynamical constraints on the orbital parameters. In coplanar case, we generate four stability maps in the $(a_b,e_b)$-plane and four stability maps in the $(e_c,e_b)$-plane for different inclinations. The initial conditions in these maps are sampled with a resolution
of $50 \times 50$, thereby producing 20000, the cases of $(a_b,e_b)$ and $(e_c,e_b)$. In the non-coplanar case, we calculated 16 stability maps in the
$\left( {\cos {i_c},\cos {i_b}}\right)$-plane for different longitudes of ascending node of Proxima b and c, generating 40,000 grid points. For a system like Proxima Centauri, the integration time should be $5.9 \times {10^2}$ - $5.9
\times {10^3}$ yr (${10^3} - {10^4}~{T_c}$) \citep{Cincotta2000,Nunez2000,Giordano2001}. In order to derive a more reliable result, we extend the integrations to $10^5$ yr for each pair of $(a_b,e_b)$, which is roughly equal to $1.698 \times {10^5} $ periods of the outer planet.

\subsection{A test case of MEGNO indicator}

As shown in Fig.5, we set $\omega_b=\omega_c ={310^\circ}$ and $M_b=M_c ={160^\circ}$ in a coplanar system.
Panel a of Fig.5 shows  that the average MEGNO of $< Y(t) >$ gradually converges to a fixed value  of  about 2 over the timescale of 1 Myr, which is reminiscent of the stable motion.
Panel b further exhibits that LCN agrees well with its estimation by $2<Y(t)>/t$. From Fig.5, we note that the estimate of LCN calculated by MEGNO converges faster to zero than that of directly integrating the variational equations. Both profiles give approximately the same slope. Hence, we can safely apply MEGNO to explore the stability of Proxima Centauri system.

\begin{figure}
\includegraphics[scale=0.35]{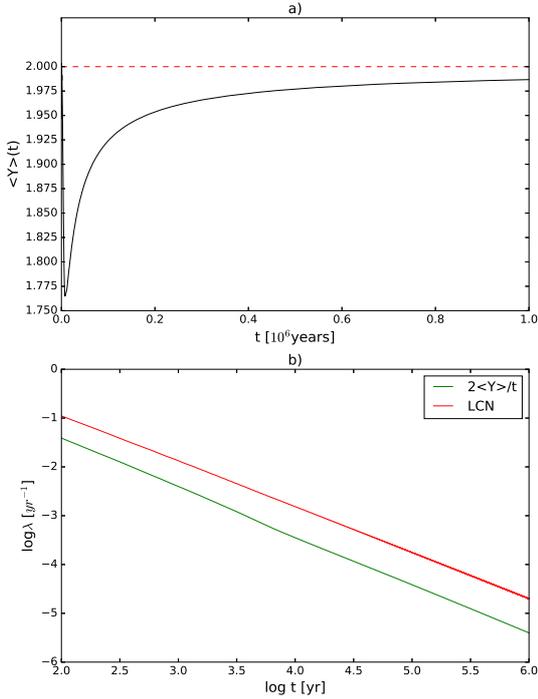}
  \centering
  \caption{The MEGNO calculated for the Proxima Centauri system. a) The mean
  $<Y(t)>$ for Proxima b over $10^6$ yr; b) the LCN computed by the direct variational method, and its
  estimation by the evolution law $2<Y(t)>/t$ of the MEGNO. The integrator accuracy is set to $10^{ - 12}$.}\label{fig5}
 \end{figure}

\begin{figure*}
 \includegraphics[scale=0.4]{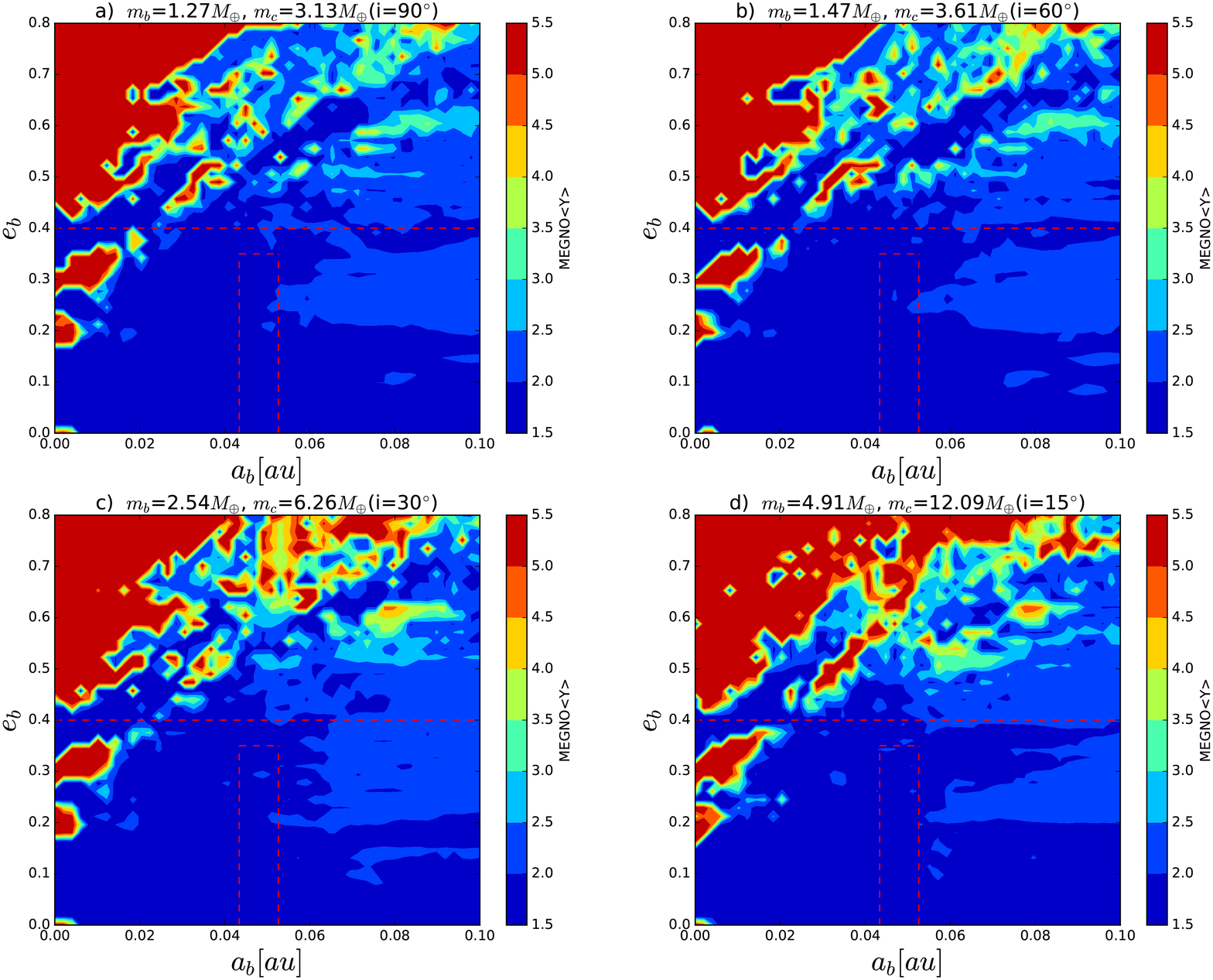}
  \centering
   \caption{Stability portraits of the Proxima Centauri system for the coplanar case in the $(a_b,e_b)$-plane. The initial
   conditions are listed in Table 1. Four panels correspond to different inclinations: a) $i = {90^{\circ} }$, b) $i =
   {60^{\circ} }$, c) $i = {30^{\circ}}$ and d) $i = {15^{\circ}}$. Each panel has a resolution
   of $50 \times 50$ for the initial data. The integration timescale for each grid point lasts $10^5$ years.
   The regions in red are referred to chaotic zones, whereas blue coloured areas, near a contour value of 2, represent quasi-periodic motion of the system. For all panels the rectangles by red dashed lines denote ${e_b} < 0.35$ and 0.0434~au $< {a_b} <$ 0.0526~au \citep{Anglada-Escud¨¦2016}.
  The red line is marked with ${e_b} = 0.4$.}
  \label{fig6}
\end{figure*}

\subsection{Stability maps for the non-coplanar case}

Fig. 6 shows the stability maps of the Proxima Centauri system in the $(a_b,e_b)$-plane for the coplanar case.
The four panels in Fig.6 correspond to different inclinations: a) $i ={90^{\circ} }$, b) $i = {60^{\circ} }$, c) $i =
{30^{\circ}}$ and d) $i = {15^{\circ}}$. From the red dashed lines marked with ${e_b} = 0.4$, we infer that the orbit of Proxima b is stable if the semi-major axis
ranges from 0.02 to 0.1 au and its eccentricity is below 0.4. In
addition, the rectangles of red dashed profiles are related to ${e_b} < 0.35$ and 0.0434~au$< {a_b}$ < 0.0526~au  \citep{Anglada-Escud¨¦2016}, being indicative of that the derived best-fitting parameters for Proxima b \citep{Anglada-Escud¨¦2016}
perfectly fall within the strong stable regime. The eccentricity of Proxima b given is well consistent with those of \citet{Anglada-Escud¨¦2016}. A large portion of the red regions where ${e_b} >0.4$ are associated with chaotic zones, apart from a few  weakly stable islands in the stability maps. This indicates that it is easy to make the planetary orbits chaotic when the orbital eccentricity is beyond the critical value in the coplanar case \citep{Gozdziewski2003a}. Moreover, \citet{Gozdziewski2003a} further indicated that the boundary of dynamical stability will be restricted to low eccentricities if the outer planet's inclination decreases in  HD 37124.
In our work, for the inner planet of Proxima b, we do not observe any clear signs that the border of stability for the eccentricities
shifts when the inner planet's inclination decreases. Proxima Centauri is about 10 times smaller in size and less massive than the Sun, therefore the effective temperature is approximately half that of the Sun, and the luminosity is only 0.17$\%$. These facts suggest that Proxima b may be in the HZ of Proxima Centauri, because its distance from the host star can be compared to Earth in the Sun's HZ when rescaling the orbits. \citet{Anglada-Escud¨¦2016} showed that Proxima b resides within conservative HZ \citep{Kopparapu2013} of Proxima Centauri, which lies between 0.0423~au and 0.0816~au. It is worth mentioning that the strong stable region in the semi-major axis ranges from 0.02 au to 0.1 au and an eccentricity for  Proxima b less than 0.4, providing  supporting evidence for  the conservative HZ of Proxima Centauri system. Furthermore,  this  conservative  HZ  \citep{Kopparapu2013} can slowly move inward by 0.1 au after $\sim$ 100 Myr, reaching the current orbit of Proxima b after $\sim$ 160 Myr, and the HZ limits for dry planets of various albedos may gradually shift inward by 0.1 au after $\sim$ 10 Myr \citep{Abe2011,Barnes2016}.

Fig. 7 shows the stability portraits of the Proxima Centauri system in the $(e_c,e_b)$-plane,
where a) $i ={90^{\circ} }$, b) $i = {60^{\circ} }$, c) $i ={30^{\circ}}$ and d) $i = {15^{\circ}}$, respectively.
We infer that robust stable regions could exist if ${e_b} < 0.45$ and ${e_c} < 0.65$.
Moreover, in the case of minimum inclination (Panel d), the stability further requires ${e_c} < 0.60$. The border of dynamical stability at ${e_b} = 0.45$  supports the best-fitting constraint of ${e_b} < 0.35$ \citep{Anglada-Escud¨¦2016}. In contrast, most regions where ${e_b} >0.45$ are linked to chaotic zones,
which contain several weakly stable zones especially when ${e_b} \simeq 0.6$. A similar estimate was reported  for other systems, e.g., the HD 37124 planetary system \citep{Gozdziewski2003a} maintaining regular orbits for ${e_c} < 0.55$, when the eccentricity of the inner planet is close to the best-fit value. As a comparison, \citet{Gozdziewski2003b} constructed stability map of the HD 12661 system in the $(e_b,e_c)$-plane, and found that the eccentricity of the outer planet can be $\simeq 0.4$ when ${e_b}$ in small values, and it cannot be larger than $\simeq 0.35$ for $e_b \simeq 0.5$.

 \begin{figure*}
\includegraphics[scale=0.45]{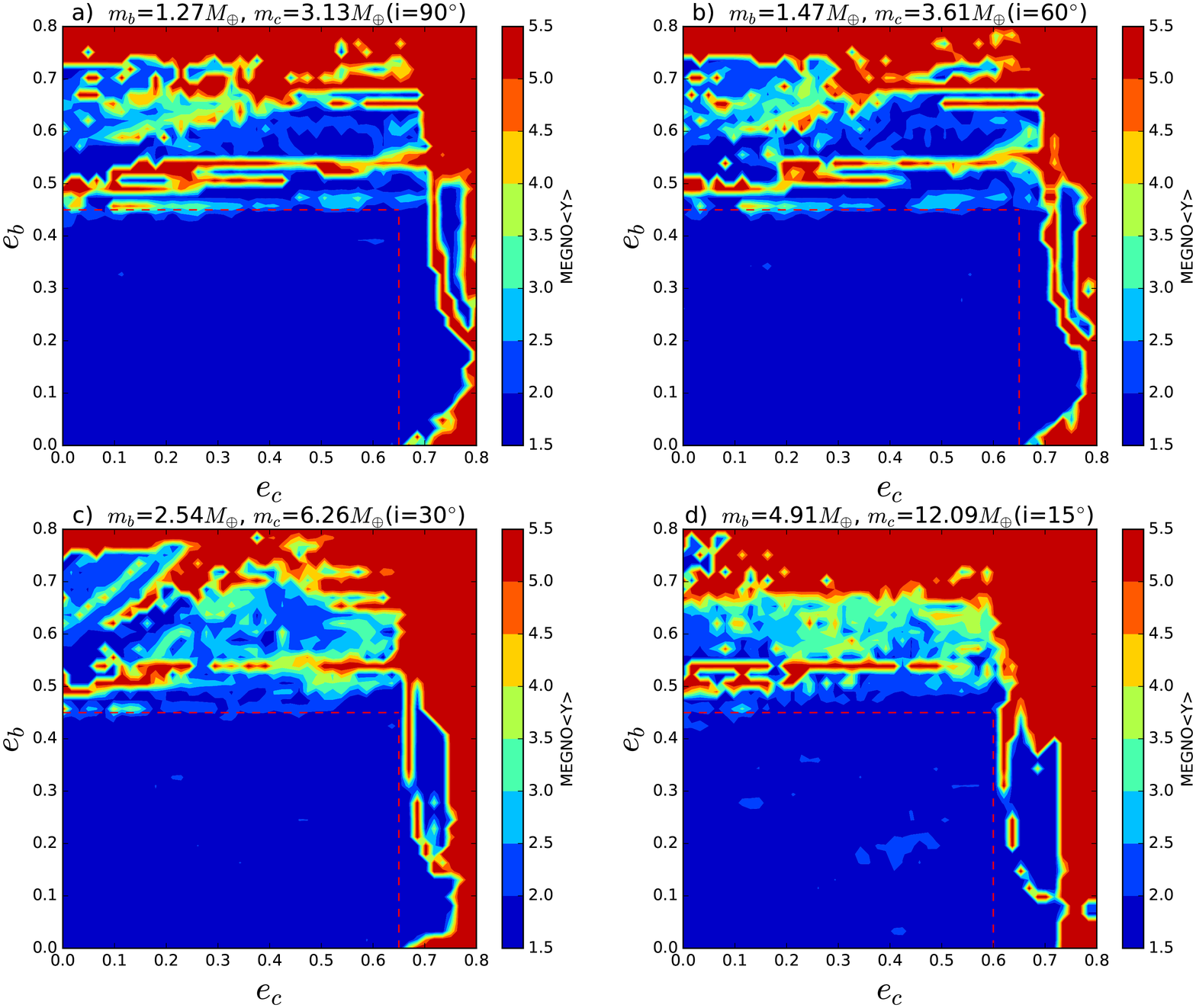}
  \centering
  \caption{In coplanar case, stability maps of the Proxima Centauri system in the $(e_c,e_b)$-plane. The initial
   parameters are from Table 1. We show four panels for different inclinations: a) $i = {90^{\circ} }$, b) $i =
  {60^{\circ} }$, c) $i = {30^{\circ}}$ and d) $i = {15^{\circ}}$. Each panel has a resolution
   of $50 \times 50$ data grids. The integration timescale  for  each data point is  $10^5$ years.
   For a), b) and c), the red dashed rectangles  represent  the region  with  ${e_b} < 0.45$ and ${e_c} < 0.65$.
   For the case of maximum mass in four cases, the rectangle of red dashed lines  represents the region with  ${e_b} < 0.45$ and ${e_c} < 0.6$. }\label{fig7}
 \end{figure*}

\subsection{Stability maps for the non-coplanar case}

Figs.8-11 show the stability maps in the $\left( {\cos {i_c},\cos {i_b}}\right)$-plane for the non-coplanar case for ${\Omega _b} = {0^{\circ} }, {90^{\circ}}, {180^{\circ}},{270^{\circ}}$, respectively.
For each ${\Omega _b}$,  we consider the variations of ${\Omega_c}$, which is adopted to be a) ${\Omega _c} = {0^{\circ} }$, b) ${\Omega _c} = {90^{\circ} }$, c) ${\Omega _c} = {180^{\circ} }$ and d) ${\Omega _c} = {270^{\circ} }$, respectively.

Left panels of Figs 8-11 show the maps of mutual orbital inclination, whereas right panels are related to the stability maps. Compared with the  maps of mutual inclination and the stability maps, the qualitative conclusion is
that the regions with higher mutual inclinations have larger MEGNO values, indicating initially higher mutual inclinations are more likely to  result in unstable planetary orbits. The red dashed lines, with respect to ${i_{mutual}} = {50^{\circ}}$, represent the current approximate border in the stability maps, between chaotic  or regular regions. We find that a great variety of initial parameters could lead to stable or quasi-periodic motions when the mutual inclinations is $\lesssim$ $50^{\circ}$. On the contrary, when the mutual inclinations exceed $50^{\circ}$, the majority of parameter space is chaotic. As a consequence, our results of mutual inclinations provide clues to potentially constraining the planetary masses of Proxima Centauri system.

For the mass of Proxima b, \citet{Brugger2017} showed that there is a 96.7$\%$ probability that the mass of Proxima b could be less than 5 $M_{\oplus}$, because the high escape velocity at the surface of a super-Earth with a mass up to 5 $M_{\oplus}$ would not hold an atmosphere. As previously mentioned, \citet{Zuluaga2018} stressed that Proxima b would be a terrestrial-mass planet of $1.3~{M_ \oplus } \le {M_p} \le 2.3~{M_ \oplus }$ where ${R_p} = 1.4_{ - 0.2}^{ + 0.3}~{R_ \oplus }$ by adopting a more elaborate evolution model of geomagnetic properties. Interestingly, the study reveals that the planet's estimated density agrees well  with the composition of a rocky planet where ${\left\langle M \right\rangle _{rocky}} = 1.63_{ -
0.72}^{ + 1.66}~{M_ \oplus }$ for its mass and ${\left\langle R \right\rangle _{rocky}} = 1.07_{ - 0.31}^{ + 0.38}~{R_ \oplus
}$ for its radius  \citep{Bixel2017}. Subsequently, on the basis of these investigations, from a dynamical viewpoint, we derive the mass range of Proxima c if ${i_{mutual}} \le {50^{\circ}}$ and $\Delta \Omega  = {\Omega _b} -{\Omega _c} = 0^{\circ}$. By varying the mass of Proxima b from a lower limit $1.27~{M_ \oplus }$ up to $1.6~{M_ \oplus }$, we conclude that the mass of Proxima c ranges from $3.13~{M_ \oplus }$ to $70.7~{M_ \oplus }$, if ${i_c}\simeq {2.5^{\circ} } - {90^{\circ} }$ and ${i_b}\simeq {52.5^{\circ} } - {90^{\circ} }$.

\begin{figure}
\includegraphics[scale=0.37]{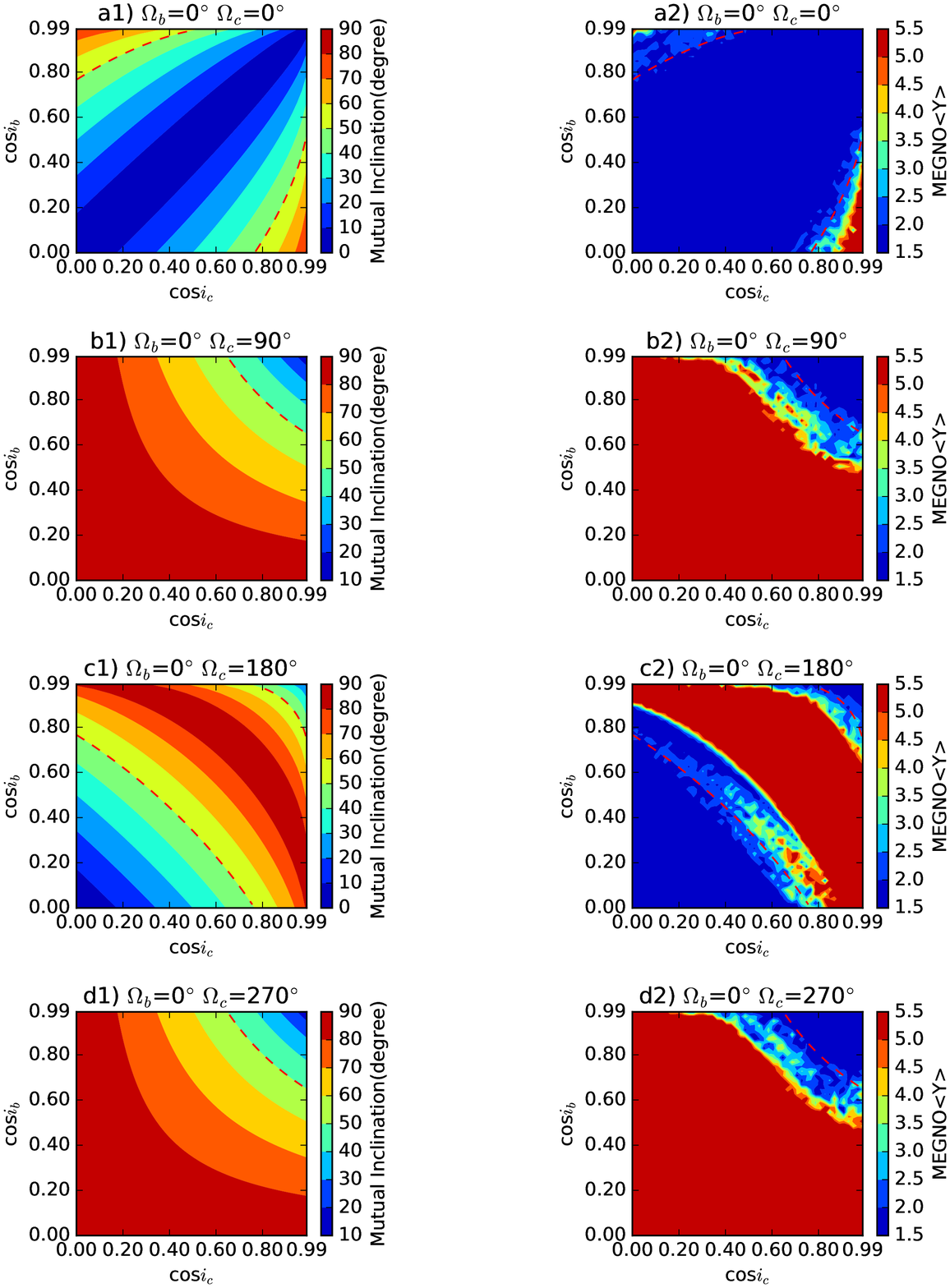}
  \centering
  \caption{Stability maps of the Proxima Centauri system in the $\left( {\cos {i_c},\cos {i_b}}
  \right)$-plane in non-coplanar case. The initial conditions are given in Table 1. Four cases for various longitude of ascending nodes: a)
  ${\Omega _b} ={0^{\circ} }, ~{\Omega _c} = {0^{\circ} }$;  b) ${\Omega _b} = {0^{\circ} },~{\Omega _c} =
  {90^{\circ} }$; c) ${\Omega _b} = {0^{\circ} },~{\Omega _c} = {180^{\circ} }$; d) ${\Omega _b} = {0^{\circ}
  },~{\Omega _c} ={270^{\circ} }$. Panel a1), b1), c1) and d1) show the maps of mutual inclination of Proxima b and
  Proxima c, respectively. Panels a2), b2), c2) and d2), respectively, present the stability maps of the system,
  which have a grid resolution of $50 \times 50$. The integration timescale for each data point is
  $10^5$ years. Red dashed curves correspond to ${i_{mutual}} = {50^{\circ}}.$ }\label{fig8}
 \end{figure}

\begin{figure}
\includegraphics[scale=0.37]{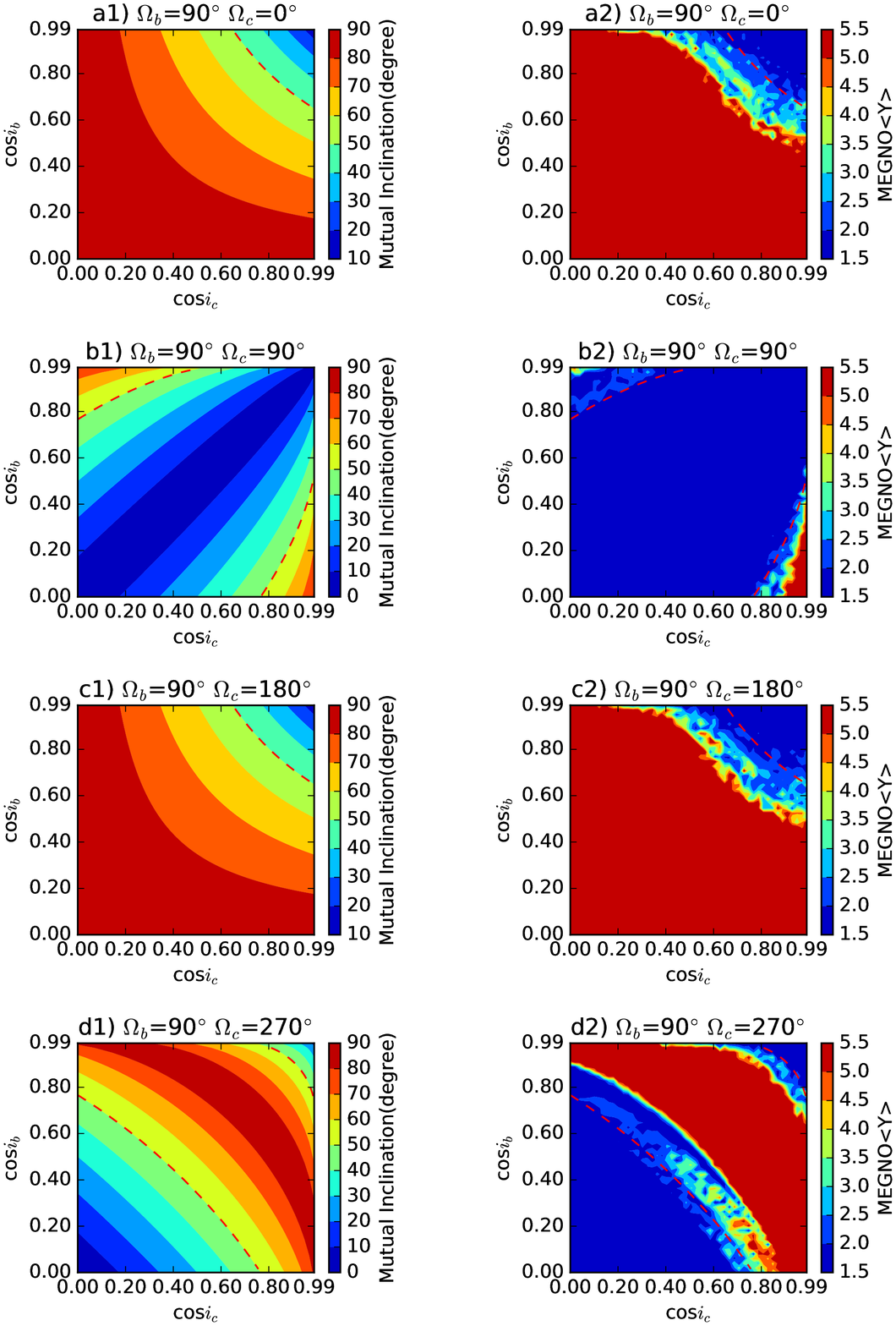}
  \centering
  \caption{In non-coplanar case, stability maps of the Proxima Centauri system in the $\left( {\cos {i_c},\cos {i_b}}
  \right)$-plane. Same as Fig.8, but for ${\Omega _b} = {90^{\circ}}$.
  }\label{fig9}
 \end{figure}

\begin{figure}
\includegraphics[scale=0.37]{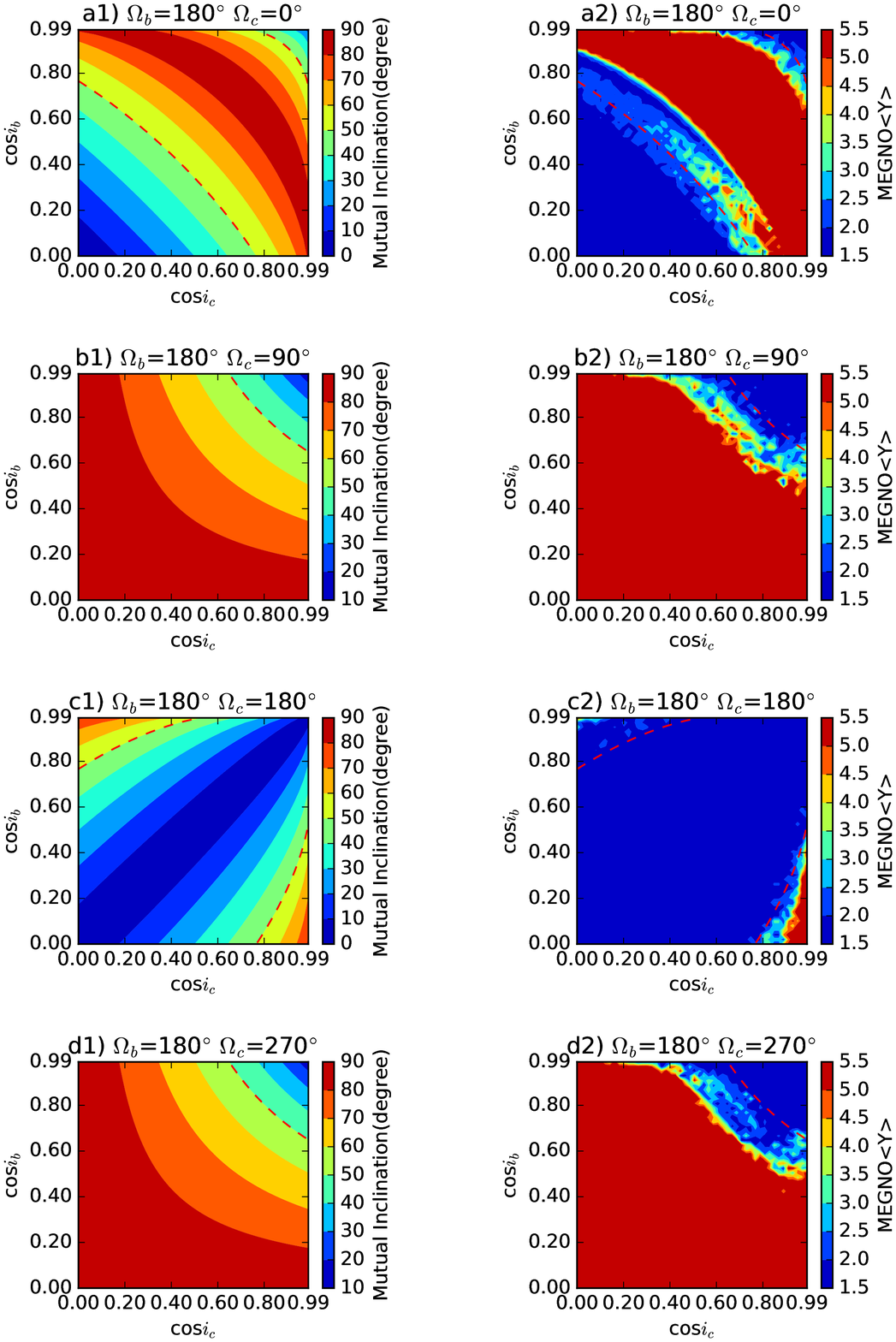}
  \centering
  \caption{In non-coplanar case, stability maps of the Proxima Centauri system in the $\left( {\cos {i_c},\cos {i_b}}
  \right)$-plane. Same as Fig.8, but for ${\Omega _b} = {180^{\circ}}$.
  }\label{fig10}
 \end{figure}

\begin{figure}
\includegraphics[scale=0.37]{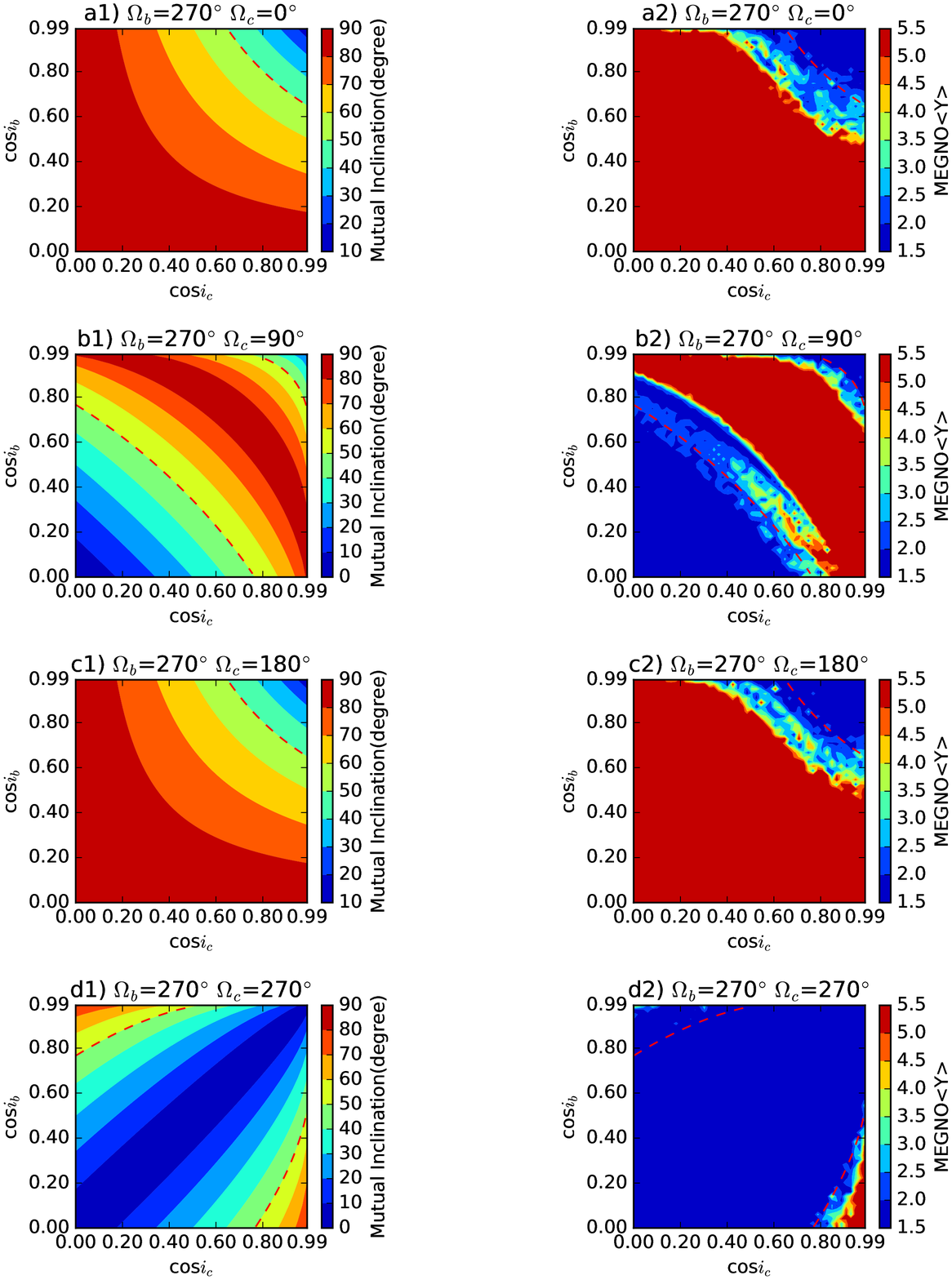}
  \centering
  \caption{In non-coplanar case, stability maps of the Proxima Centauri system in the $\left( {\cos {i_c},\cos {i_b}}
  \right)$-plane.  Same as Fig.8,  but for ${\Omega _b} = {270^{\circ}}$.
  }\label{fig11}
 \end{figure}

\section{Conclusions and discussion}

In this work, we extensively explore the dynamical evolution and stability of Proxima Centauri system, which consists of an Earth-mass planet (Proxima b) and a candidate planet (Proxima c) with an orbital period of approximately 215 days \citep{Anglada-Escud¨¦2016,Barnes2016}. For the coplanar case, we study the evolution of orbital eccentricities of the Proxima Centauri system by numerical integration of the full equations of motion and with a semi-analytical model that accounts for relativistic and tidal effects.
For coplanar and non-coplanar configurations, we further estimate the dynamical limits on orbital parameters that provide stable or quasi-periodic motions of the Proxima Centauri system, using the MEGNO technique. Subsequently, we constrain the approximate mass range for Proxima c based on dynamical constraints from the simulations. Herein, We summarize the principal results as follows.

For the coplanar case, both the numerical integrations and the semi-analytical model show that the relativistic effect plays a major role in the evolution of eccentricities of two planetary orbits, whereas the tidal effect only has an influence on the eccentricity of Proxima b over a long timescale.
In addition, we compare the eccentricity evolution outcomes for two planets from the modified secular equations with those from direct integrations, and we find good mutual agreement.

Secondly, we show stability maps in the $(a_b,e_b)$-plane for the coplanar case coplanar system,
where strong stable regions of Proxima b exist for the coplanar case with the semi-major axis ranging from 0.02 au to 0.1 au and the eccentricity being less than 0.4. However, most regions where ${e_b} >0.4$ are chaotic. \citet{Anglada-Escud¨¦2016} emphasized that Proxima b resides within the conservative HZ of Proxima Centauri 0.0423 - 0.0816 au \citep{Kopparapu2013}. The best-fitting parameters for Proxima b \citep{Anglada-Escud¨¦2016} fall exactly within this stable regime, providing evidence that the Proxima Centauri system lies within the HZ. Moreover, we extensively investigate the stability in the $(e_c,e_b)$-plane for the coplanar case. We conclude that robust stability of this system would require ${e_b} < 0.45$ and ${e_c} < 0.65$.

Thirdly, we explore different initial values for longitudes of ascending node of $0^{\circ}$, $90^{\circ}$, $180^{\circ}$ and $270^{\circ}$ in the non-coplanar systems, then we investigate the stability maps of the Proxima Centauri system in the $\left( {\cos {i_c},\cos {i_b}}\right)$-plane. Systems with higher mutual inclinations are more likely to be chaotic. We conclude that mutual inclinations lower than $50^{\circ}$ could regularise the Proxima Centauri system. Moreover, we estimate the mass of Proxima c assuming that ${i_{mutual}} \le {50^{\circ}}$ and $\Delta \Omega  = {\Omega _b} -{\Omega _c} = 0^{\circ}$, and suggest the mass of Proxima c may range from $3.13~{M_\oplus }$ to $70.7~{M_ \oplus }$ with respect to ${i_c}\simeq {2.5^{\circ} } - {90^{\circ} }$, if the mass of Proxima b ranges from $1.27~{M_\oplus }$ up to $1.6~{M_ \oplus }$. Our work gives sound dynamical arguments in favour of the physical feasibility of the system, thereby motivating further observational efforts to confirm or rule out the presence of Proxima c in the future.

\section*{Acknowledgments}

We thank the referees for constructive comments and suggestions.
We acknowledge Javier Mart{\'i} and Cristi{\'a}n Beaug{\'e} for their discussions.
We appreciate Y.X. Gong and S. Wang for suggestions on the manuscript.
This work is financially supported by the
National Natural Science Foundation of China (Grants No. 11773081,
11573073, 11873097), CAS Interdisciplinary Innovation Team,
the Foundation of Minor Planets of the Purple Mountain Observatory.

\label{lastpage}

\end{document}